\documentclass[aps,prl,twocolumn,superscriptaddress,showpacs]{revtex4}
\usepackage{graphicx,color}
\usepackage{amsmath}
\usepackage{amssymb}
\usepackage{bm}

\def\uparrow{\delimiter"0222378 }
\def\downarrow{\delimiter"0223379 }

\begin{document}
\title{Entanglement Generation by a Three-Dimensional Qubit Scattering:\\Concurrence vs.\ Path (In)Distinguishability}

\author{Yuichiro Hida}
\affiliation{Department of Physics, Waseda University, Tokyo
169-8555, Japan}

\author{Hiromichi Nakazato}
\affiliation{Department of Physics, Waseda University, Tokyo
169-8555, Japan}

\author{Kazuya Yuasa}
\affiliation{Waseda Institute for Advanced Study, Waseda University,
Tokyo 169-8050, Japan}

\author{Yasser Omar}
\affiliation{CEMAPRE, ISEG, Universidade T\'{e}cnica de Lisboa, P-1200-781 Lisbon, and 
SQIG, Instituto de Telecomunica\c{c}\~oes, P-1049-001 Lisbon, Portugal}

\date[]{April 30, 2009}

\begin{abstract}
A scheme for generating an entangled state in a two spin-$1/2$ system by means of a spin-dependent potential scattering of another qubit is presented and analyzed in three dimensions.
The entanglement is evaluated in terms of the concurrence both at the lowest and in full order in perturbation with an appropriate renormalization for the latter, and its characteristics are discussed in the context of (in)distinguishability of alternative paths for a quantum particle. 
\end{abstract}

\maketitle

Entanglement is one of the most peculiar features of quantum theory with no classical analog and plays an essential role in quantum information and technology \cite{ref:schemes}, though its acquisition or controlled generation is by no means a trivial matter. 
There are proposals for its generation and one often makes use of their mutual interaction to let the two quantum systems entangled \cite{ref:QdotTheories}.  
On the other hand, when they are far apart and/or their mutual interaction is (in theory for all practical purposes) absent, one may resort to another quantum system (``mediator'') to make the two parties entangled, through the individual and successive local interactions of the former with the latters \cite{ref:1Dcases,ref:Haroche,ref:Kuzmich,ref:1Dsc}.
This kind of scheme has been investigated for simple systems of qubits (quantum two-level systems), usually under the assumption that the strength of the interaction between the mediator and each qubit is completely under our control \cite{ref:1Dcases,ref:Haroche}.

Even though the assumption is considered to be legitimate, for example, when the interaction is well controlled by switching on/off the external parameters \cite{ref:Haroche}, there are still cases in which it is untenable or its applicability is questionable.
In particular, when the interaction time is not well defined or its definition necessarily requires a resolution in some conceptual issues, like the moments of the beginning and the end of the interaction for a particle described by a wave packet with a finite width and scattered by a static potential, we would be forced to treat the process as a quantum mechanical scattering process of a mediator system off the target, where additional (internal, e.g., spin) degrees of freedom are duly taken into account.
In the scattering problem, the interaction strength is in a sense automatically and implicitly given and we have no choice of defining or controlling the interaction duration once the initial conditions have been fixed.
It is therefore an interesting and nontrivial matter of physical relevance to examine whether the schemes for entanglement generation or extraction based on the interaction between the mediator and the subsystems could remain valid and function properly even when one has little controllability on such parameters as time.

In this article, a three-dimensional scattering process, in which a mediator spin-$1/2$ qubit is scattered off a fixed target composed of two other spin-$1/2$ qubits by spin-flipping $\delta$-shaped potentials, is considered to examine the ability of obtaining an entangled state in the target system. 
The same setup has already been considered, but essentially only in one (spatial) dimension, to generate a maximally entangled state in the two-qubit system by tuning the interaction strengths \cite{ref:1Dcases} or, as a one-dimensional scattering process, by adjusting the incident momentum of the mediator (or the distance between two target qubits) \cite{ref:1Dsc}.

Notice that the treatments of the scattering processes so far are not considered to be completely satisfactory because in one dimension, there would be no way to incorporate such important physical parameters as the incident and scattering angles, the size of the wave packet and the detector resolution.
The purpose here is to take these elements into account in the scattering process and to show that they actually rule the resulting entanglement. 
We shall also see an interesting connection between the concurrence and the path (in)distinguishability of the particles.

\begin{figure}
\includegraphics[width=0.3\textwidth]{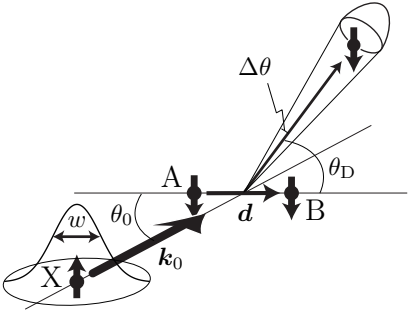}
\caption{Qubit X, with wave-packet width $w$ and central momentum $\bm{k}_0$, is sent toward qubits A and B (fixed at positions $-\bm{d}/2$ and $\bm{d}/2$, respectively) with an incident angle $\theta_0$. X is then scattered by A and B with an angle $\theta_\text{D}$ (measured from $\bm d$), and detected by a detector with a finite resolution characterized by an opening angle $\Delta\theta$.}
\label{fig:Setup}
\end{figure}

We want to generate entanglement between two spin-$1/2$ qubits A and B, initially in the separable state $|\downarrow\downarrow\rangle_\text{AB}$.
Assume that these qubits are fixed at positions $-{\bm d}/2$ and ${\bm d}/2$, and that there is no interaction between them.
In order to make them entangled, a third spin-$1/2$ qubit X is prepared in the up state $|\uparrow\rangle_\text{X}$ and sent toward the system A+B\@.
X is then scattered by spin-flipping $\delta$-shaped potentials produced by A and B and is finally detected by a spin-sensitive detector with a finite opening, as described in Fig.\ \ref{fig:Setup}.
This physical process can be described by the total Hamiltonian
\begin{align}
H=\frac{{\bm p}^2}{2m}
&{}+g\delta^3({\bm r}+{\bm d}/2)
\bigl({\bm\sigma}^\text{(X)}\cdot{\bm\sigma}^\text{(A)}\bigr)
\nonumber\\
&{}+g\delta^3({\bm r}-{\bm d}/2)
\bigl({\bm\sigma}^\text{(X)}\cdot{\bm\sigma}^\text{(B)}\bigr),
\label{eq:H}
\end{align}
where $\bm p$ and $\bm r$ are the momentum and the position of qubit X of mass $m$, ${\bm\sigma}^\text{(J)}$ ($\text{J}=\text{X}, \text{A}, \text{B}$) the Pauli matrices acting on qubit J, and $g$ the coupling constant.
We shall treat two qubits A and B symmetrically, for simplicity.

Since the Hamiltonian $H$ preserves the total number of ups ($\uparrow$) among the three qubits X, A, and B, if we find X in the spin-flipped (i.e., down) state at the detector after scattering, we are sure that one of the qubits, A \textit{or} B, must be in a spin-flipped state, that is, system A+B is either in $|\uparrow\downarrow\rangle_\text{AB}$ \textit{or} in $|\downarrow\uparrow\rangle_\text{AB}$.
If there is no way to judge which spin has been flipped in the scattering process with X, the qubits A and B are certainly in their superposed state, i.e., an entangled state.
Notice, however, that the state finally to be extracted can no longer be a pure state, because it shall be given by a reduced density matrix after being traced over possible spatial or momentum degrees of freedom of X within the detector resolution and therefore it becomes mixed in general.
We assume adiabatic switchings of interaction and no (abrupt) changes of the parameters in the Hamiltonian are considered to occur.   
As it is well known, the quantity of physical relevance in the scattering problem is the element of the scattering matrix (S) which describes the transition from the remote past $t\to-\infty$ to the remote future $t\to+\infty$ under the Hamiltonian $H$.

Let the incident qubit X be described by a Gaussian wave packet, with a central momentum ${\bm k}_0$ and a (spatial) width $w$, so that the initial state of the total system reads as
$|\psi_0\rangle
=\int d^3{\bm k}\,\psi_0 ({\bm k})
|{\bm k}\uparrow\downarrow\downarrow\rangle_\text{XAB}$, 
$\psi_0 ({\bm k})
=(2w^2/\pi)^{3/4}e^{-w^2({\bm k}-{\bm k}_0)^2}$.
The spin-sensitive detector detects X scattered in the direction ${\hat{\bm n}}_\text{D}$ seen from the origin (scattering center) with an opening angle $\Delta\theta$.
When X has been found in $|\downarrow\rangle_\text{X}$ in the detector that covers the solid angle $\Delta\Omega=2\pi(1-\cos\Delta\theta)$
around direction $\hat{\bm n}_\text{D}$, the state of the target system A+B is given by the reduced density matrix of the form
\begin{align}
\rho
={}&P^{-1}\int_{\Delta\Omega}d^3{\bm k}\,
{}_\text{X}\hspace*{-0.2mm}\langle{\bm k}\downarrow|S|\psi_0\rangle
\langle\psi_0|S^\dagger|{\bm k}\downarrow\rangle_\text{X}
\nonumber\displaybreak[0]\\
={}&
P^{-1}\,\Bigl(
a_{11}|\uparrow\downarrow\rangle_\text{AB}
\langle\uparrow\downarrow|
+a_{22}|\downarrow\uparrow\rangle_\text{AB}
\langle\downarrow\uparrow|
\nonumber\displaybreak[0]\\
&\qquad\quad
{}+a_{12}|\uparrow\downarrow\rangle_\text{AB}
\langle\downarrow\uparrow|
+a_{12}^*|\downarrow\uparrow\rangle_\text{AB}
\langle\uparrow\downarrow|
\Bigr),
\label{eq:rhoAB}
\end{align}
where $S$ denotes the S matrix. 
The normalization constant $P=a_{11}+a_{22}$ 
is nothing but the probability that this particular event occurs, i.e., the yield.
Given the state $\rho$ for two qubits, the degree of entanglement can be measured in terms of its concurrence $C(\rho)$ \cite{ref:concurrence}, which reads, for the above $\rho$ in (\ref{eq:rhoAB}), as
\begin{equation}
C(\rho)=\frac{2|a_{12}|}{a_{11}+a_{22}}.
\end{equation}

The S-matrix element is given by 
$\langle{\bm k}'\zeta'|S|{\bm k}\zeta\rangle=\delta^3({\bm k}'-{\bm k})\delta_{\zeta'\zeta}-2\pi i\delta(E_{k'}-E_k)\langle{\bm k}'\zeta'|V|\Psi_{\bm k}\zeta\rangle$, 
where $|{\bm k}\zeta\rangle$ is the eigenstate of the free Hamiltonian $H_0={\bm p}^2/2m$ and $|\Psi_{\bm k}\zeta\rangle$ that of the total Hamiltonian $H=H_0+V$, both belonging to the same eigenvalue $E_k=\hbar^2{\bm k}^2/2m$, with $\zeta$ denoting the spin degrees of freedom.
The coordinate representation of the latter reads as
$\langle{\bm r}|\Psi_{\bm k}\zeta\rangle=\langle{\bm r}|{\bm k}\zeta\rangle-\int d^3{\bm r}'\,G_k({\bm r}-{\bm r}')\frac{2m}{\hbar^2}V({\bm r}')\langle{\bm r}'|\Psi_{\bm k}\zeta\rangle
$, where $G_k(\bm{r})$ is the retarded Green function.

Up to the first order in $g$ (Born approximation), the relevant S-matrix elements are given by (subscripts ${}_\text{X,A,B}$ for spin states shall be suppressed) 
\begin{align}
&\langle{\bm k}\downarrow\uparrow\downarrow|S|\psi_0\rangle
=-\langle{\bm k}\downarrow\downarrow\uparrow|S|\psi_0\rangle^*
\equiv A({\bm k})
\nonumber\\
&\ \ %
=-\frac{ig}{2\pi^2}\int d^3{\bm k}'\,\psi_0 ({\bm k}')\delta(E_k-E_{k'})e^{i({\bm k}-{\bm k}')\cdot{\bm d}/2}.
\label{eq:A(k)}
\end{align}
The matrix elements $a_{ij}$ are simply expressed as $a_{11}=a_{22}=\int_{\Delta\Omega}d^3{\bm k}\,|A({\bm k})|^2$ and $a_{12}=-\int_{\Delta\Omega}d^3{\bm k}\,A^2({\bm k})$.
When the incident wave packet $\psi_0({\bm k})$ is well monochromatized, $wk_0\gg1$, $A({\bm k})$ is approximately evaluated analytically and we end up with the following expressions for the concurrence and the yield at the lowest order
\begin{gather}
C(\rho)
\simeq\frac{1}{\Delta\Omega}
\left|\int_{\Delta\Omega}d^2\hat{\bm k}\, 
e^{ik_0(\hat{\bm k}\cdot{\bm d})-[(\hat{\bm k}-{\hat{\bm k}}_0)\cdot{\bm d}]^2/8w^2}\right|,
\label{eq:CBorn}
\\
P\simeq\frac{m^2g^2}{\pi^3\hbar^4w^2}\,\Delta\Omega\,e^{-[{\bm d}^2-({\hat{\bm k}}_0\cdot{\bm d})^2]/8w^2}.
\label{eq:PBorn}
\end{gather}

\begin{figure}
\includegraphics[width=0.48\textwidth]{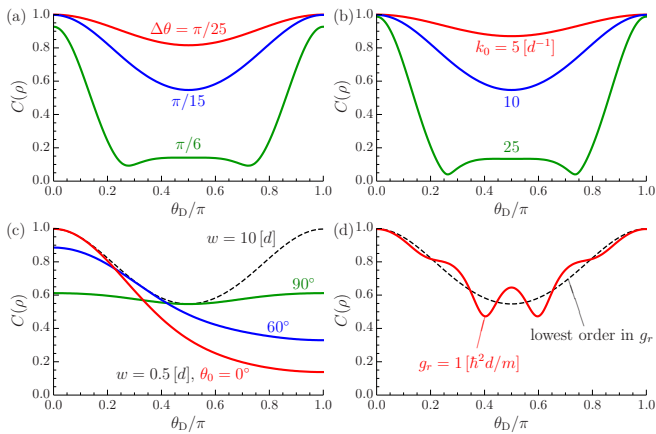}
\caption{(Color online) Concurrence $C(\rho)$ as a function of the scattering (or detecting) angle $\theta_\text{D}$, (a)--(c) evaluated at the lowest and (d) in full order (solid line) in perturbation.
(a),(b),(d) are for $w\gg d$ and (c) is for $w=0.5d$.
The other parameters are (if not specified in the figures): $\theta_0=\pi/2$, $\Delta\theta=\pi/15$, $k_0d=10$, $g_r'=0$.}
\label{fig:C}
\end{figure}

The angle integrations over $\hat{\bm k}$ in the concurrence $C(\rho)$ in (\ref{eq:CBorn}) are numerically performed and we find the following characteristics of it [Figs.\ \ref{fig:C}(a)--(c)].
When the incident qubit X is accompanied with a large wave packet $w\gg d$ [Figs.\ \ref{fig:C}(a)--(b)], 
(i) the concurrence $C(\rho)$ depends on in which direction the scattered qubit X is detected and (ii) takes the maximal value $\sim1$ when X is captured in the direction $\pm{\bm d}$, i.e., on the line connecting target qubits A and B, while (iii) no apparent dependence is seen on the incident angle $\theta_0$. 
(iv) It is reduced considerably as $\Delta\theta$ is increased, while it keeps the maximal value in the $\pm{\bm d}$ directions [Fig.\ \ref{fig:C}(a)].  
Furthermore, (v) it strongly depends on $k_0$ and reduces considerably for a large $k_0$ or a short wavelength compared with the distance $d$ between the two qubits A and B in the target [Fig.\ \ref{fig:C}(b)].
If the incident wave packet is small compared with the distance between two qubits A and B, that is, $w\lesssim d$, the concurrence $C(\rho)$ becomes deteriorated in general and the incident-angle dependence, which is almost absent in the opposite cases with $w\gg d$, appears [Fig.\ \ref{fig:C}(c)].  
We observe that (vi) there are directions where the concurrence still takes the same value as that in the case of $w\gg d$, i.e., when X is detected in the same direction as the incident direction $\hat{\bm k}_0$ and in its symmetric direction with respect to ${\bm d}$. 
As for the probability $P$ in (\ref{eq:PBorn}), it does not depend on scattering (detection) angle $\theta_\text{D}$ and Young-like interference is not observed.

The characteristics of the concurrence $C(\rho)$ can be understood on the basis of its mathematical expression in (\ref{eq:CBorn}) or its approximate expression \cite{ref:fullpaper}
\begin{equation}
C(\rho)\sim1-\frac{1}{8}(k_0d)^2(\Delta\theta)^2\sin^2\!\theta_\text{D},
 \,\text{for $w\gg d$ and $\Delta\theta\ll1$}.
\end{equation}
We understand that this expression indeed well describes the characteristics (i)--(v) mentioned above.
The condition for obtaining a higher concurrence is also attainable by evaluating the range of the variation of the phase of the integrand in (\ref{eq:CBorn}) over the opening of the detector mouth, which should be smaller than $2\pi$,
\begin{equation}
2k_0d\sin\Delta\theta\sin\theta_\text{D}\lesssim2\pi,\quad\text{for $w\gg d$}.
\label{eq:highClw}
\end{equation}
On the other hand, if the incident wave packet is small $w\lesssim d$, we have to keep the second term $-[(\hat{\bm k}-\hat{\bm k}_0)\cdot{\bm d}]^2/8w^2$ in the exponent in (\ref{eq:CBorn}).
Since this term would entail an exponential reduction of $C(\rho)$, the condition for keeping a higher concurrence becomes
\begin{equation}
|(\hat{\bm k}-\hat{\bm k}_0)\cdot{\bm d}|\ll w,
\quad\text{for $w\lesssim d$},
\label{eq:highCsw}
\end{equation}
which explains well the characteristics  (vi).

It would be interesting to interpret the above conditions (\ref{eq:highClw}) and (\ref{eq:highCsw}) for higher entanglement, in the context of the (in)distinguishability of the paths taken by particle X\@.
As a general rule in quantum theory, the (in)distinguishability of alternative paths of a particle results in the (non)vanishing of quantum interference.
Since the concurrence $C(\rho)$ is proportional to the absolute value of the off-diagonal matrix element $a_{12}$, its value is rather dependent on the information about which qubit A or B has changed its spin state in the scattering process.
Notice also that in the lowest-order perturbation, since qubit X with its spin flipped is scattered only and surely once, either by qubit A or B, and the interaction certainly changes their spin states, the information about which qubit has scattered X is the same as that about which spin has flipped.
That is why the probability $P$ does not exhibit Young-like interference: the path taken by X can \textit{in principle} be disclosed by checking the spin state of A and B after the scattering. This distinguishability erases the interference.
The same,  applies to the concurrence.
If one could distinguish the two alternative paths of X, originating from A or B, one is able to know which spin has been flipped.
This knowledge results in a reduction of the off-diagonal matrix elements and therefore of the concurrence $C(\rho)$.

We understand that the conditions for higher concurrence (entanglement) (\ref{eq:highClw}) and (\ref{eq:highCsw}) would be interpreted as those for the indistinguishability of the two alternative paths from A or B\@.
Indeed, if the incident wave packet is long $w\gg d$ and therefore is approximately considered as a plane wave of wavelength $\lambda_0=2\pi/k_0$, the condition for higher concurrence (\ref{eq:highClw}) may be understood in comparison with the resolving power of an optical device. 
It is known in classical optics that the optical device that has an aperture $\Delta\theta$ seen from an object composed of two optical sources with mutual distance $d$ is unable to distinguish the two sources if the condition 
$\Delta\theta\lesssim\lambda_0/d\sin\theta_\text{D}$
is satisfied (for $\Delta\theta\ll1$).  
This is essentially the same as the condition (\ref{eq:highClw}).
For the opposite case with $w\lesssim d$, since the quantity $|(\hat{\bm k}-\hat{\bm k}_0)\cdot{\bm d}|$ is nothing but the difference in length between the two paths via A or B, if the condition (\ref{eq:highCsw}) is not satisfied, one could determine the path the particle X has passed through on its way to the detector, for the path length difference is certainly larger than the size of the particle $w$.
In both cases, the conditions for higher entanglement are interpreted as those for indistinguishability of particle paths.

Notice that this interpretation and the connection with the indistinguishability are limited to the lowest-order results, since in higher order in perturbation, multiple scatterings that invalidate the one to one correspondence between the knowledge of the particle paths and that of the spin flips come into play.  
One also realizes that higher-order terms require a proper treatment, for otherwise the result would become trivial, that is, no scattering could occur by the $\delta$-shaped potentials in (spatial) dimensions greater than one.
This is a famous issue in quantum theory, and Jackiw has proposed a prescription how to deal with such systems.
According to Jackiw \cite{ref:Jackiw}, we have to renormalize the strength of the $\delta$-shaped potential so that the source term becomes nonvanishing.
The coupling constant in the Hamiltonian, $g$ in our case, has to be regarded as a bare one and absorb possible divergences arising from the Green function at the origin $G_k({\bm0})$.

Furthermore, in the present case, another element that was absent at that time, i.e., the spin degrees of freedom, turns out to require another care in dealing with higher-order terms.
In this respect, it is important to notice that the interaction of the form $g{\bm\sigma}^\text{(X)}\cdot{\bm\sigma}^\text{(A)}$ inevitably causes another type of interaction in its higher-order terms.
For example, in its second order, a term proportional to unit operator in spin space, that is not proportional to the original form, appears,
$
(g{\bm\sigma}^\text{(X)}\cdot{\bm\sigma}^\text{(A)})^2
=3g^2-2g^2{\bm\sigma}^\text{(X)}\cdot{\bm\sigma}^\text{(A)}.
$
The situation is more clearly understood from the fact that the interaction can be written as ${\bm\sigma}^\text{(X)}\cdot{\bm\sigma}^\text{(A)}=\mathcal{P}_\text{3XA}-3\mathcal{P}_\text{1XA}$ in terms of projection operators on the spin-triplet and singlet spaces, $\mathcal{P}_\text{3XA}$ and $\mathcal{P}_\text{1XA}$.
This means that in any higher-order terms in perturbation, there are only two types of interactions proportional to $\mathcal{P}_\text{3XA}$ or $\mathcal{P}_\text{1XA}$ present and we need to renormalize these two terms simultaneously.
That is, we need two counter terms to obtain sensible results.
We therefore introduce another bare coupling constant $g'$, start with a bare interaction Hamiltonian $V({\bm r})=
\delta^3({\bm r}+{\bm d}/2)(g{\bm\sigma}^\text{(X)}\cdot{\bm\sigma}^\text{(A)}+g')
+\delta^3({\bm r}-{\bm d}/2)(g{\bm\sigma}^\text{(X)}\cdot{\bm\sigma}^\text{(B)}+g')$, and renormalize the bare coupling constants
\begin{subequations}
\begin{gather}
\frac{1}{g_r+g_r'}=\frac{1}{g+g'}+\frac{2m}{\hbar^2}\lim_{\Lambda\to\infty}\frac{1}{2\pi^2}\Lambda
,\displaybreak[0]\\
\frac{1}{-3g_r+g_r'}=\frac{1}{-3g+g'}+\frac{2m}{\hbar^2}\lim_{\Lambda\to\infty}\frac{1}{2\pi^2}\Lambda,
\end{gather}
\end{subequations}
where the linear divergence is due to the divergent part of $G_k({\bm0})$ in three dimensions.
After the renormalization, everything is finite and nonvanishing and we obtain the relevant S-matrix elements in full order, which can be evaluated numerically and even analytically for the case of large wave packet $w\gg d$, and the concurrence \cite{ref:fullpaper}.

Figure \ref{fig:C}(d) shows the concurrence as a function of the scattering (detection) angle $\theta_\text{D}$, evaluated both at the lowest order (Born approximation) and in full order in perturbation.
We observe here that the characteristics of the concurrence at the lowest order mentioned above  are largely maintained in full order even though the effect of multiple scatterings is manifest as an appearance of wrinkles in the central part.
This implies that when qubit X is detected in some particular directions, one can expect a higher concurrence or entanglement produced in the target qubits A and B owing to the multiple scatterings, though the optimal one is still obtainable in the direction parallel to $\bm d$ (i.e., $\theta_\text{D}=0,\,\pi$).

We have studied a scheme for the entanglement generation between two fixed spins by scattering a mediator spin in three dimension. 
The close connection between the concurrence, a measure of entanglement, and the path (in)distinguishability, is disclosed, which seems to be valid in a global sense even in higher orders in perturbation. 
It is remarkable that we would not need to control the interaction time to obtain the higher entanglement and this possibility may strengthen the way of entanglement generation by means of quantum scatterings.
The details of calculations and further analysis will be reported elsewhere \cite{ref:fullpaper}.

This work is partly supported by a Grant-in-Aid
for Scientific Research (C) from JSPS, Japan, by a Special Coordination Fund for Promoting Science and Technology from MEXT, Japan, by the the bilateral Italian-Japanese Projects of  MUR, Italy, and by the Joint Italian-Japanese Laboratory 
of MAE, Italy. 
Y. Omar thanks the support from Funda\c{c}\~{a}o para a Ci\^{e}ncia e a
Tecnologia (Portugal), namely through programs POCTI/POCI/PTDC, partially funded by FEDER (EU).

\end{document}